\documentclass[12pt,a4paper]{article}
\usepackage[format=hang]{caption}
\usepackage{epsfig,amssymb,amsmath,graphicx,subcaption,verbatim,xcolor,ulem,epstopdf}
\usepackage{graphics,color}
\newcommand{\be}{\begin{equation}}
\newcommand{\ee}{\end{equation}}

\newcommand{\news}{\setcounter{equation}{0}}
\def\bea{\begin{eqnarray}}
\def\eea{\end{eqnarray}}
\numberwithin{equation}{section}

\begin{document} 

\title{\vskip -70pt
\begin{flushright}
{\normalsize DAMTP-2014-58} \\
\end{flushright}
\vskip 60pt
{\bf {\LARGE A Skyrme model approach to the spin-orbit force}}\\[20pt]}
\author{\bf {\Large C. J. Halcrow\footnote{C.J.Halcrow@damtp.cam.ac.uk}\, and N.S. Manton\footnote{N.S.Manton@damtp.cam.ac.uk}} \\[25pt]
Department of Applied Mathematics and Theoretical Physics\\
University of Cambridge\\
Wilberforce Road, Cambridge CB3 0WA, UK}

\date{October 2014}
\maketitle
\vskip 40pt
 
\begin{abstract}
The spin-orbit force is a vital tool in describing finite nuclei and nucleon interactions; however its microscopic origin is not fully understood. In this paper we study a model inspired by Skyrmions which provides a classical explanation of the force. To simplify the calculations the Skyrmions are approximated as two-dimensional rigid discs which behave like quantum cogwheels.
\end{abstract}

\vskip 80pt

Keywords: Skyrmions, Spin-orbit force
\newpage



\section{Introduction}\news

The spin-orbit force is a crucial ingredient in many parts of nuclear physics \cite{Greiner}. In the elementary shell model, nuclei are described as a collection of nucleons which do not directly interact. They only interact indirectly through an effective potential which gives rise to a one-particle Hamiltonian and consequently an energy spectrum. By the Pauli exclusion principle, levels of this energy spectrum are filled as the baryon number $B$ is increased. For special values of $B$, the spectrum hits a gap and the corresponding nucleus is tightly bound and very stable. These special values are called magic numbers and give rise to magic nuclei. The shell model works well near these. To obtain the correct magic numbers one must include a spin-orbit term in the single particle Hamiltonian \cite{Otto,Mayer}. This couples the spin of a nucleon to its orbital angular momentum $l$. The inclusion of this term breaks the degeneracy between states with the same value of $|l|$. States with spin and orbital angular momentum aligned are energetically favoured. 

For a nucleus with a few more nucleons than a magic number we can interpret its structure physically: a core made from a magic nucleus is surrounded by the other nucleons orbiting it. If we have one orbiting nucleon, its spin and orbital angular momentum are aligned in all but two cases, Antimony-$133$ and Bismuth-$209$. As more nucleons are added, other factors such as pairing make the interpretation more complicated. The spin-orbit force is strongest near the surface of the core and its physical meaning is lost within the core.

Analogy with atomic physics points to an electromagnetic origin of the spin-orbit coupling but this turns out to have the wrong magnitude. The correct magnitude can be obtained by considering relativistic effects. They lead to a field theory where nucleons interact via mesons. The system can be solved approximately by neglecting quantum fluctuations of certain terms \cite{Ring}. While this technique is successful, it ignores the structure of nucleons and requires one to fit several parameters. Ideally these parameters would come from experiment but as the theory is phenomenological, effective masses and coupling constants must be used \cite{Ring2}.

The spin-orbit force is also present in nucleon-nucleon interactions. It couples the orbital angular momentum to the sum of the spins of the nucleons, and can also be thought of as coming from meson interactions. The asymptotic form of the force has been successfully reproduced in the Skyrme model using a product ansatz which is valid only at large separations \cite{Riska} .

In this paper we develop an idea in \cite{MSkySD} which provides an explanation for the spin-orbit force at shorter separations, inspired by the Skyrme model. We introduce the Skyrme model in section $2$, describing the important features which shape the spin-orbit interaction. We then set up a precise, simplified model of Skyrmion-Skyrmion interactions and solve it in section $3$, first for the simpler case of nucleon-nucleon interactions and then for the case of a nucleon interacting with a larger nucleus, which describes certain shell model configurations.

\section{The Skyrme Model}

The Skyrme model is a nonlinear field theory of pions which admits soliton solutions called Skyrmions \cite{Sk}. These are identified as nuclei with the topological charge $B$ of a Skyrmion equal to the baryon number of the system.

A solution is generally represented by a surface of constant baryon density which is coloured to express the direction of the pion field as it varies over the surface; we use the same colouring scheme as in \cite{108}. The spherically symmetric $B=1$ solution, known as the hedgehog, is displayed in figure $1$. When two $B=1$ Skyrmions are widely separated we can approximate their interaction using an asymptotic expansion. One finds that among all configurations there is a special submanifold of maximal attraction between the Skyrmions called the attractive channel \cite{book}. This is easiest to interpret pictorially: in the attractive channel the separated Skyrmions have matching colours at the point of closest contact. Conversely, if the closest colours are opposite the Skyrmions repel.

\begin{figure}[ht]
	\centering
	\includegraphics[width=4cm]{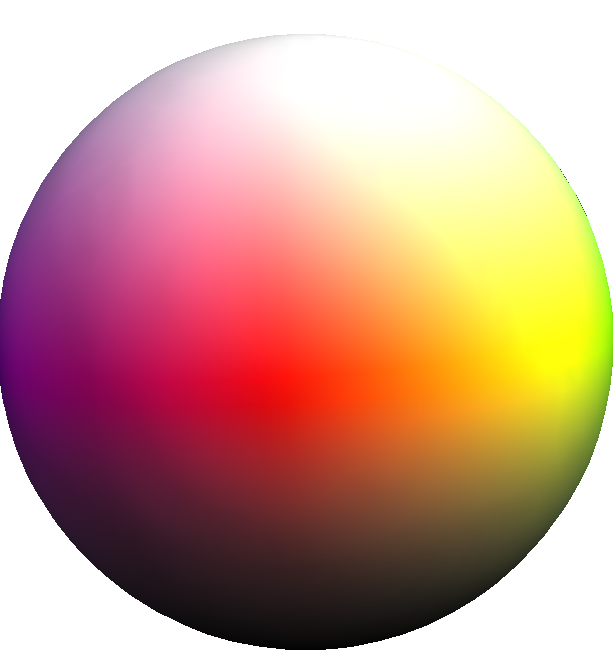}
	\caption{The $B=1$ Skyrmion solution.}
	\label{fig:B1}
\end{figure}

\begin{figure}[ht]
	\centering
	\includegraphics[width=8cm]{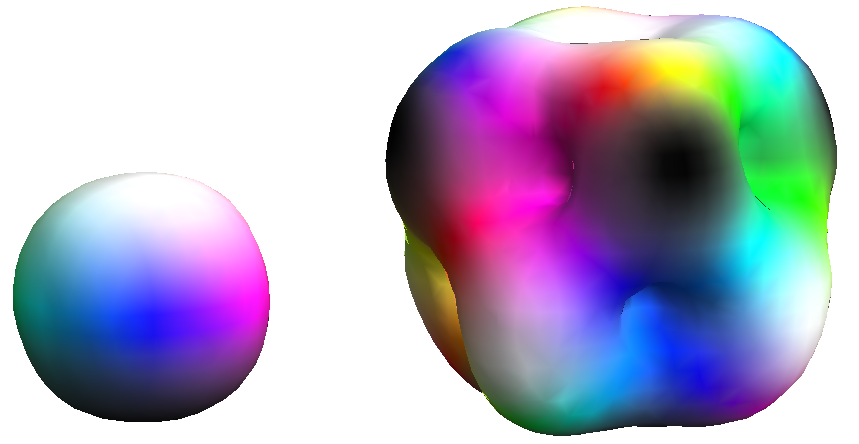}
	\caption{A $B=1$ Skyrmion close to a $B=6$ Skyrmion. The colours of closest contact are both red (unseen on the $B=1$ solution from this viewpoint) so the configuration is in the attractive channel.}
	\label{fig:B1B6}
\end{figure}

Configurations tend to line up in the attractive channel in order to minimise potential energy. This concept remains useful for larger Skyrmions. As an example, consider the configuration in figure \ref{fig:B1B6}. Here, a $B=1$ Skyrmion is orbiting a $B=6$ Skyrmion. The system is shown in the attractive channel with red on both Skyrmions at their contact point. To stay in the attractive channel as it orbits, the $B=1$ Skyrmion must take a special orbital path. Specifically, it rolls around the equator of the larger solution completing three full rotations on its axis before returning to the initial position. The key observation is that the $B=1$ Skyrmions' orbital angular momentum is aligned with its spin. This is exactly what is required for the spin-orbit force in nuclei with $B$ one more than a magic number, except in the cases of Antimony-$133$ and Bismuth-$209$. It is the classical pion field structure of Skyrmions that provides the microscopic origin for the coupling. Many other Skyrmion pairs have paths like this which encourage spin-orbit coupling. The effect becomes stronger when the Skyrmions are closer together but loses meaning if they were to merge fully. This is consistent with the fact that the traditional spin-orbit force is strongest near the surface of the core nucleus. 

We will now try to work out the consequences of this classical spin-orbit coupling when the system is quantised. The usual procedure for quantising one Skyrmion is to use a rigid body approach to the classical minimal energy configuration, promoting its collective coordinates to quantum operators. Quantising the interaction between separated Skyrmions is more difficult and little progress has been possible using the full set of collective coordinates \cite{Manifolds}. Thus, we will only consider a toy model in two dimensions where we treat the Skyrmions as rigid discs. We will begin by carefully considering the simplest system possible: the interaction of two $B=1$ Skyrmions.

\section{Discs interacting through a contact potential}

\subsection{Two discs of equal size}

\begin{figure}[ht]
\centering
\includegraphics[width=15cm]{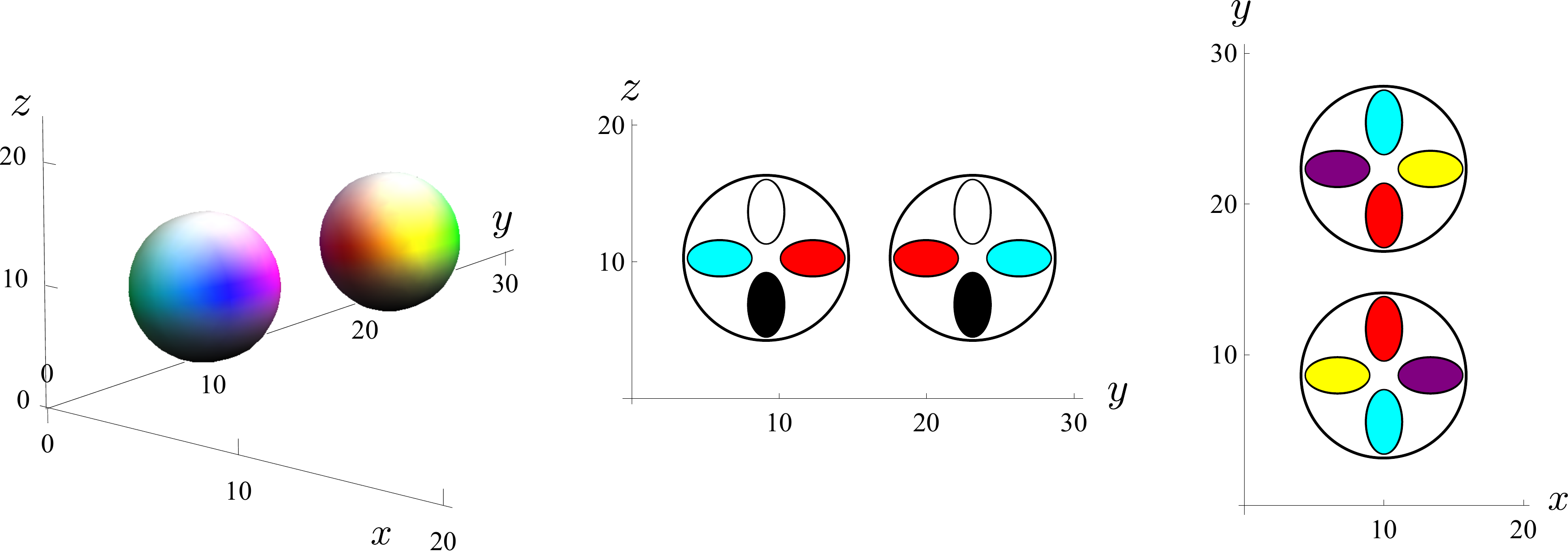}
\caption{(a) Two $B=1$ Skyrmions in the attractive channel. (b) The rolling configuration. (c) The sliding configuration}
\label{fig:2Dslice}
\end{figure}

Our model is based on taking $2$D slices of $3$D Skyrmion configurations, taking our inspiration from $B=1$ Skyrmion interactions. Figure \ref{fig:2Dslice}a shows separated $B=1$ solutions in the attractive channel. We can take a $2$D slice of this parallel to the $y$-$z$ plane and parallel to the $x$-$y$ plane to give us the systems in figures \ref{fig:2Dslice}b and \ref{fig:2Dslice}c. We now treat these $2$D objects as rigid discs, at fixed separation, interacting through a potential which depends only on their colouring. 

\begin{figure}[ht]
	\centering
	\includegraphics[width=6cm]{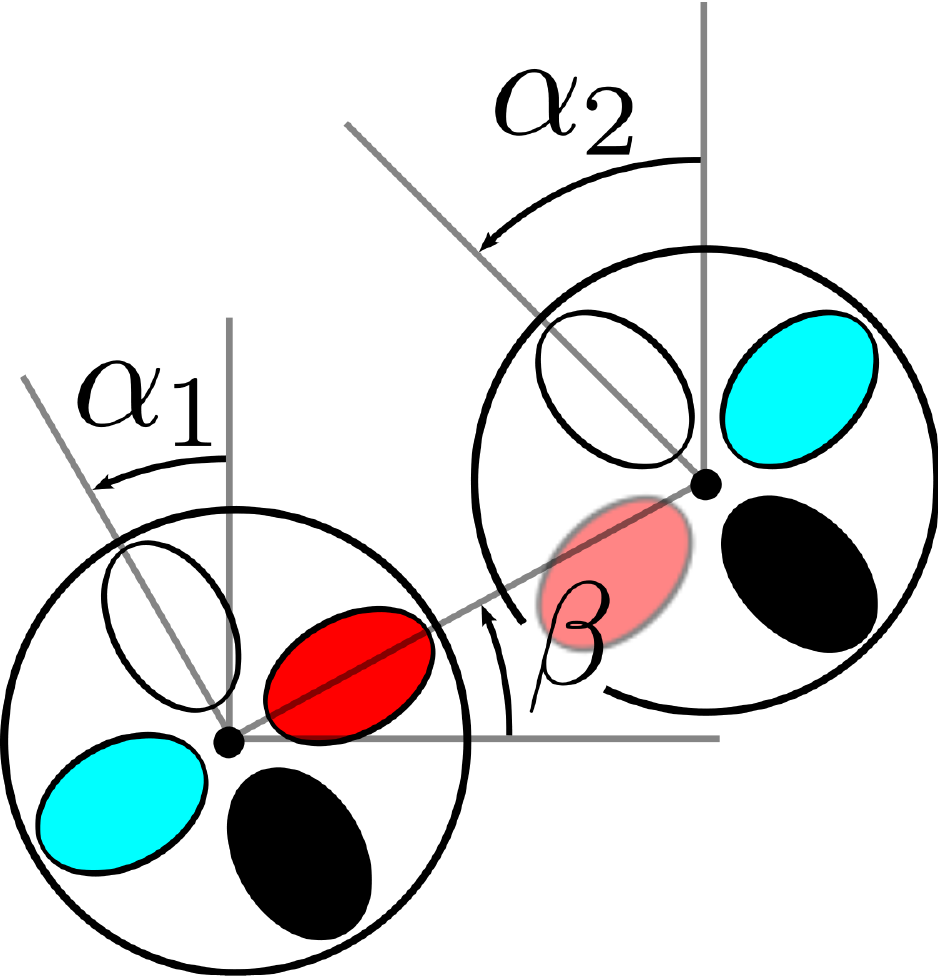}
	\caption{The angles $\alpha_1$, $\alpha_2$ and $\beta$.}
	\label{fig:angles}
\end{figure}

To remain in the attractive channel the discs in figures \ref{fig:2Dslice}b and  \ref{fig:2Dslice}c must, respectively, roll and slide around each other. For now, we will consider the rolling configuration. Labelling the discs as $1$ and $2$ we introduce the angular coordinates as in figure \ref{fig:angles}. The angles $\alpha_1$ and $\alpha_2$ represent the orientation of the discs with respect to their own axes. These are measured anti-clockwise and are zero when white points up, as in figure \ref{fig:2Dslice}b. The coordinate $\beta$ labels the orbital orientation of the discs while $r$ is the (fixed) distance between the disc centres. 

Each of the coordinates has range $2\pi$ but $\beta \to \beta + \pi$ also returns the system to the attractive channel. As such, the potential must be periodic under full rotations of either disc and under half an orbital rotation. It should also only depend on the colouring: the simplest choice is a cosine potential. Thus, a classical Lagrangian which describes the system is
\begin{equation}
L = \frac{1}{2}I_1\dot{\alpha}_1^2 + \frac{1}{2}I_2\dot{\alpha}_2^2 + \frac{1}{2}\mu r^2\dot{\beta}^2+ k\cos\left(2\beta - \alpha_1 - \alpha_2\right) 
\end{equation}
where $I_1$, $I_2$ are the disc moments of inertia, $\mu$ is the reduced mass of the system and $k>0$ is the strength of the potential. The argument of the potential measures the difference in colour at the closest points. The discs are identical so $I_1=I_2=I$ and we may write $\mu r^2$ in terms of $I$ by introducing a dimensionless separation parameter $d$ and setting $\mu r^2 = 4d^2I$. This simplifies the Lagrangian to
\begin{equation} \label{lag}
L = \frac{1}{2}I\left(\dot{\alpha}_1^2 + \dot{\alpha}_2^2 + 4d^2\dot{\beta}^2\right)+ k\cos\left(2\beta - \alpha_1 - \alpha_2\right) \,.
\end{equation}

Classically, the lowest energy solution satisfies $2\beta - \alpha_1 - \alpha_2 = 0$. This forces the discs into the attractive channel as if they were cogwheels; the first cog rolls around the second, fixed cog. If they stay in the attractive channel for all time, we can differentiate this condition to obtain a relation between velocities: $2\dot{\beta} - \dot{\alpha}_1 - \dot{\alpha}_2 = 0$. Introducing the classical conjugate momenta to the coordinates
\begin{equation}
 s_1 = I\dot{\alpha}_1\,,\, s_2 = I\dot{\alpha}_2 \text{ and } l = 4d^2I\dot{\beta}\,,
 \end{equation}
we can rewrite the above velocity relation as
\begin{equation} \label{cog}
l - 2d^2(s_1+s_2) = 0\,.
\end{equation}
Later we will see that this combination of spins and angular momentum has an important role to play in the quantum picture too.

The Lagrangian \eqref{lag} has two linearly independent continuous symmetries. The first corresponds to all angles increasing by the same amount. This leads to conservation of total angular momentum
\begin{equation} \label{Jspin}
\mathcal{J} = I\left(\dot{\alpha}_1+\dot{\alpha}_2+4d^2\dot{\beta}\right) = s_1 + s_2 + l\,.
\end{equation}
The other conserved quantity is generated by one disc spinning at the same speed as the other but in the opposite direction. Since this quantity can be interpreted purely in terms of the colour fields moving, we label it as the total isospin in analogy with the full Skyrme model. It has the form
\begin{equation} \label{Ispin}
\mathcal{I} = I\left(\dot{\alpha}_1-\dot{\alpha}_2\right) = s_1-s_2\,.
\end{equation}

We can take advantage of these symmetries by changing coordinates and reducing the problem's degrees of freedom from three to one. Before doing this, we should consider the discrete symmetries of the system which occur since the configuration space is a $3$-torus. First let us solve the problem for $k=0$ where the Hamiltonian becomes that of a free particle on a $3$-torus. After canonical quantisation, the Hamiltonian has the form
\begin{equation}
\hat{\mathcal{H}} = -\frac{1}{2I}\left( \frac{\partial^2}{\partial\alpha_1^2}+\frac{\partial^2}{\partial\alpha_2^2}+\frac{1}{4d^2}\frac{\partial^2}{\partial\beta^2}  \right)
\end{equation}
where we have set $\hbar = 1$. The wavefunction has the form
\begin{equation} \label{sss}
\psi_{\text{free}}\left(\alpha_1,\alpha_2,\beta\right) = e^{i(s_1\alpha_1 + s_2\alpha_2 + l\beta)}
\end{equation}
with corresponding energy
\begin{equation}
E_{\text{free}} = \frac{1}{2I}\left( s_1^2 + s_2^2 + \frac{1}{4d^2}l^2 \right)\,.
\end{equation}

The quantities $s_1$, $s_2$ and $l$ are the quantum numbers corresponding to the spins and orbital angular momentum of the free discs. As we are modelling Skyrmions, the discs are treated as fermions. Thus, the wavefunction picks up a minus sign under full disc rotations: $\alpha_1 \to \alpha_1 + 2\pi$ and $\alpha_2 \to \alpha_2 + 2\pi$. This means $s_1$ and $s_2$ are both half-integers. The system is also invariant under $\beta \to \beta + 2\pi$ and as such $l$ must be an integer. While these quantities do not remain good quantum numbers when the potential is turned on, they do remain important due to Bloch's theorem. This says that there exists a basis of energy eigenstates of the form
\begin{equation} \label{Bloch}
	\psi\left(\alpha_1,\alpha_2,\beta\right) = e^{i(s_1\alpha_1 + s_2\alpha_2 + l\beta)}u\left(\alpha_1,\alpha_2,\beta\right)\,,
\end{equation}
where $u$ is periodic on the $3$-torus, and has the same periodicity as the potential. This theorem is generally used in an infinite lattice but we are on a torus. As such $s_1$, $s_2$ and $l$ have discrete allowed values instead of continuous ones. They are also usually defined up to a vector in the reciprocal lattice, a discrete lattice in $3$D. However we fix their value by insisting that
\begin{equation}
u\left(\alpha_1, \alpha_2, \beta\right)|_{k=0} \equiv 1\,.
\end{equation}
There is one state per cell in the reciprocal lattice. Thus we can understand $s_1$, $s_2$ and $l$ as labelling a particular lattice cell. We will see later that energy states from different cells do not cross when the potential is turned on and as such these labels are good for tracking the energy states as $k$ increases. 

To make progress we must now change coordinates to take advantage of the continuous symmetries from earlier. We introduce new coordinates $(\gamma,\xi,\eta)$. Two of these should give rise to the conjugate momenta corresponding to $\mathcal{J}$ and $\mathcal{I}$. That is
\begin{align}
-i\frac{\partial}{\partial\xi} &= -i\left(\frac{\partial}{\partial\alpha_1} + \frac{\partial}{\partial\alpha_2}+\frac{\partial}{\partial\beta}\right)\\
-i\frac{\partial}{\partial\eta} &= -i\left(\frac{\partial}{\partial\alpha_1} - \frac{\partial}{\partial\alpha_2}\right) \,.
\end{align}
Note that these operators commute with the potential in \eqref{lag}. We may define $\gamma$ to be the coordinate in the potential. If we also insist on a diagonal quadratic kinetic term in the Hamiltonian we arrive at a unique coordinate transformation
\begin{equation} \label{changeo}
\begin{pmatrix} \gamma \\ \xi \\ \eta \end{pmatrix} = \begin{pmatrix} -1 & -1 & 2 \\ \frac{1}{2+4d^2} & \frac{1}{2 + 4d^2} & \frac{4d^2}{2+4d^2} \\ \frac{1}{2} & -\frac{1}{2} & 0 \end{pmatrix} \begin{pmatrix} \alpha_1 \\ \alpha_2 \\ \beta \end{pmatrix} \,.
\end{equation}
This transforms the Hamiltonian to
\begin{equation}
\hat{\mathcal{H}} = -\frac{1}{2I}\left(\frac{1+2d^2}{d^2}\frac{\partial^2}{\partial\gamma^2} + \frac{1}{2+4d^2}\frac{\partial^2}{\partial\xi^2} + \frac{1}{2}\frac{\partial^2}{\partial\eta^2}\right) - k\cos\gamma\,.
\end{equation}

Since the $\xi$ and $\eta$ contributions are purely kinetic, the wavefunction has the form
\begin{equation}
\psi(\gamma,\eta,\xi) = e^{i\mathcal{J}\xi}e^{i\mathcal{I}\eta}\chi(\gamma)\,.
\end{equation}
Moreover, after applying the coordinate transformation, comparison with \eqref{sss} and \eqref{Bloch} tells us that
\begin{align}
\mathcal{J} &= s_1 + s_2 + l\,, \\
\mathcal{I} &= s_1 - s_2
\end{align}
as in the classical equations \eqref{Jspin} and \eqref{Ispin}, and that
\begin{equation}
\chi(\gamma) = e^{iq_\gamma \gamma}\tilde{u}(\gamma)\,,
\end{equation}
where
\begin{equation}
q_\gamma = \frac{1}{2+4d^2}\left(l - 2d^2(s_1 + s_2)\right)
\end{equation}
and $\tilde{u}(\gamma)$ has period $2\pi$. Once again, we fix $q_\gamma$ so that $\tilde{u}|_{k=0} \equiv 1$. From earlier, we find that $\mathcal{I}$ and $\mathcal{J}$ can take any integer values. The free system now has the wavefunction
\begin{equation}
\psi_{\text{free}}(\gamma,\eta,\xi) = e^{i\mathcal{J}\xi}e^{i\mathcal{I}\eta}e^{iq_\gamma\gamma}
\end{equation}
with corresponding energy
\begin{equation}
E_{\text{free}} = \frac{1+2d^2}{2d^2I}q_\gamma^2 + \frac{1}{(4 + 8d^2)I}\mathcal{J}^2 + \frac{1}{4I}\mathcal{I}^2\,.
\end{equation}
For fixed $\mathcal{J}$ and $\mathcal{I}$, the allowed values of $q_\gamma$ are separated by integers, though the fractional part of $q_\gamma$ depends on $d$ and $\mathcal{J}$. Combining everything, the problem reduces to the Schr\"{o}dinger equation
\begin{align} \label{schro}
-\frac{1+2d^2}{2d^2I}\frac{d^2 }{d \gamma^2}\left( e^{iq_\gamma \gamma}\tilde{u}\right) - k\cos\gamma \, e^{iq_\gamma \gamma}\tilde{u} &= \left( E - \frac{\mathcal{J}^2}{(4+8d^2)I} - \frac{\mathcal{I}^2}{4I}  \right) e^{iq_\gamma \gamma}\tilde{u}\\
&\equiv E_\gamma e^{iq_\gamma \gamma}\tilde{u}\,.
\end{align}

This is the Mathieu equation, which has been extensively studied \cite{mathi}. We will now consider it with our physical picture in mind. The energy has separated into two parts -- one depends on $\mathcal{J}$ and $\mathcal{I}$ and has no $k$ dependence. The other only depends on the $\gamma$ sector. The potential does not mix states with different $\mathcal{I}$ and $\mathcal{J}$. Thus, we can fix these values and focus on calculating $E_\gamma$.

We can understand the system when $k$ is small by using perturbation theory. Note that the dimensionless small quantity is really $kI$. The energy, to second order in $kI$ is
\begin{equation} \label{perteng}
E_{\gamma,\text{pert}} =  \frac{1+2d^2}{2d^2I}q_\gamma^2 + (kI)^2\frac{d^2}{(1+2d^2)I}\frac{1}{4q_\gamma^2 -1}\,.
\end{equation}
The most important thing to note is that for fixed $\mathcal{J}$ and $\mathcal{I}$, since the allowed values of $q_\gamma$ are separated by integers, there is a unique state which satisfies $4q_\gamma^2 - 1 < 0$. Thus there is one state whose energy decreases after perturbation, with $|q_\gamma| \leq \frac{1}{2}$. We call states which satisfy this condition energetically favourable. At $|q_\gamma| = \frac{1}{2}$ equation \eqref{perteng} breaks down and degenerate perturbation theory must be used.  It tells us that the energy spectrum develops a gap at each of these points leaving a separated energy band for $|q_\gamma| < \frac{1}{2}$ which does not touch the rest of the spectrum. 

\begin{figure}[ht]
	\centering
	\includegraphics[width=12cm]{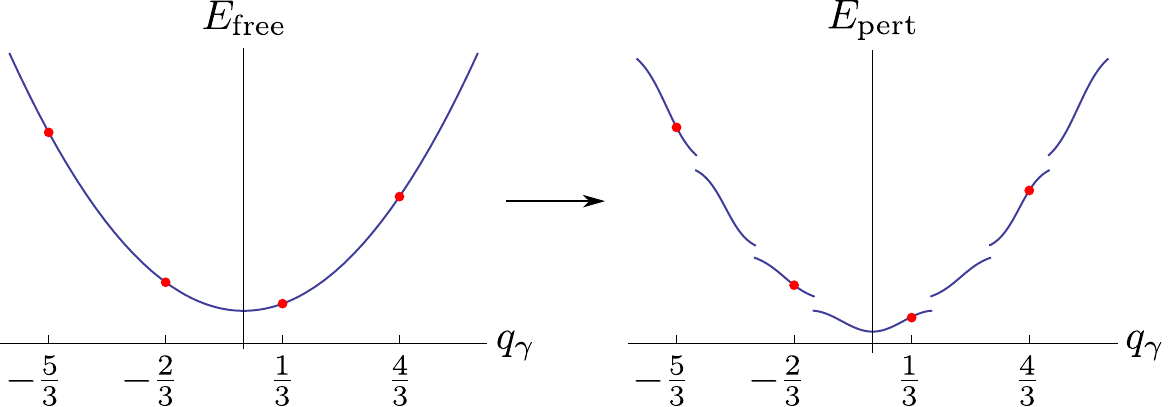}
	\caption{How the energy spectrum changes after perturbation. $E_{\text{free}}$ is the spectrum for $k=0$; $E_{\text{pert}}$ is the spectrum for small $kI$. The dots represent an example of an allowed value of $q_\gamma$. In this case we take $(\mathcal{I},\mathcal{J}) = (0,1)$ and $d = 1$ which gives $q_\gamma \equiv \frac{1}{3} (\text{mod } 1)$. Note that there is one allowed state per separated band.}
	\label{fig:pereng}
\end{figure}

The other degenerate points ($|q_\gamma| = 1, \frac{3}{2}, 2$...) lead to singularities in the perturbative energy spectrum at higher orders. Degenerate perturbation theory tells us, once again, that a gap occurs at each of these points. Thus after perturbation we are left with an energy spectrum divided into non-touching bands as seen in figure \ref{fig:pereng}. Degenerate points are identified and as such each band is an integer long. For example, one of the bands is $q_\gamma \in [-1,-\frac{1}{2}] \cup [\frac{1}{2},1]$. Since the allowed values of $q_\gamma$ are separated by an integer there is exactly one state per band. This explains why $q_\gamma$ is a good label: due to the gaps in the spectrum we can follow a free state as $k$ increases without having to worry about crossing except at degenerate points. Even there, the uncertainty is only between two states and most degeneracies only occur for special values of $d$. As such, we won't consider them carefully.

For large $k$ we may use a tight binding (\text{tb}) limit. This approximation relies on the wavefunction being concentrated within each unit cell in $\gamma$ with negligible overlap. Then the total wavefunction can be written as a sum of isolated wavefunctions which solve Schr\"odinger's equation within the unit cell. These isolated wavefunctions must be the same at each site due to the periodicity of $\tilde{u}$. Bloch's theorem allows for the total wavefunction to pick up a phase between cells meaning the solution of \eqref{schro} is of the form
\begin{equation}
e^{iq_\gamma\gamma}\tilde{u}_{\text{tb}}(\gamma) = \sum_{m \in \mathbb{Z}}  \phi(\gamma - 2\pi m) e^{2\pi iLm}
\end{equation}
where $\phi$ is the isolated wavefunction and $L$ is some constant. The periodicity of $\tilde{u}$ fixes $L$ to be $q_\gamma$. Thus our total, tight binding wavefunction is
\begin{equation}
\psi_{\text{tb}}(\gamma, \xi, \eta) = e^{i\mathcal{J}\xi}e^{i\mathcal{I}\eta}\sum_{m\in \mathbb{Z}}\phi(\gamma - 2\pi m)e^{2\pi i mq_\gamma}\,.
\end{equation}

We are left to find $\phi$. Since $k$ is large, we assume that the wavefunction is concentrated near the minimum of the potential. We can expand the potential near this point, which gives
\begin{equation} \label{pot}
-k\cos\gamma \approx -k\left(1 - \frac{\gamma^2}{2} + \frac{\gamma^4}{4!}\right)
\end{equation}
and reduces the Schr\"odinger equation \eqref{schro} to
\begin{equation} \label{phischro}
-\frac{1+2d^2}{2d^2I}\frac{d^2\phi}{d\gamma^2} - k\left(1 - \frac{\gamma^2}{2} + \frac{\gamma^4}{24}\right)\phi = E_\gamma\phi\,,
\end{equation}
where $kI$ is large. If we temporarily ignore the $\gamma^4$ term in the potential then this truncated Schr\"odinger equation is just a simple harmonic oscillator which can be solved by standard methods. The non-normalised eigenstates are given by
\begin{equation}
\phi_N(\gamma) = H_N\left(\left(\frac{kd^2I}{1+2d^2}\right)^\frac{1}{4}\gamma\right)\exp\left(-\left(\frac{kd^2I}{4+8d^2}\right)^\frac{1}{2}\gamma^2\right)
\end{equation}
where $H_N$ are the Hermite polynomials. We can then use these to find the energies to $O(1)$ in $k$. They are
\begin{equation} \label{endo}
E_{\gamma,n} = -k + \sqrt{k}\sqrt{\frac{1+2d^2}{d^2I}}\left(N+\frac{1}{2}\right) - \frac{1+2d^2}{32d^2I}\left(2N^2 + 2N + 1\right) + O\left( \frac{1}{ \sqrt{k} } \right)\,.
\end{equation}
The $O(k)$ term is from the constant in the potential. The $O(\sqrt{k})$ term is the usual harmonic oscillator energy, and the $O(1)$ term is the contribution from the $\gamma^4$ term in the potential, evaluated by first order perturbation theory. We have ignored all overlap terms between cells, but these are exponentially suppressed for large enough $k$. 

\begin{figure}[ht]
	\centering
	\includegraphics[width=13.8cm]{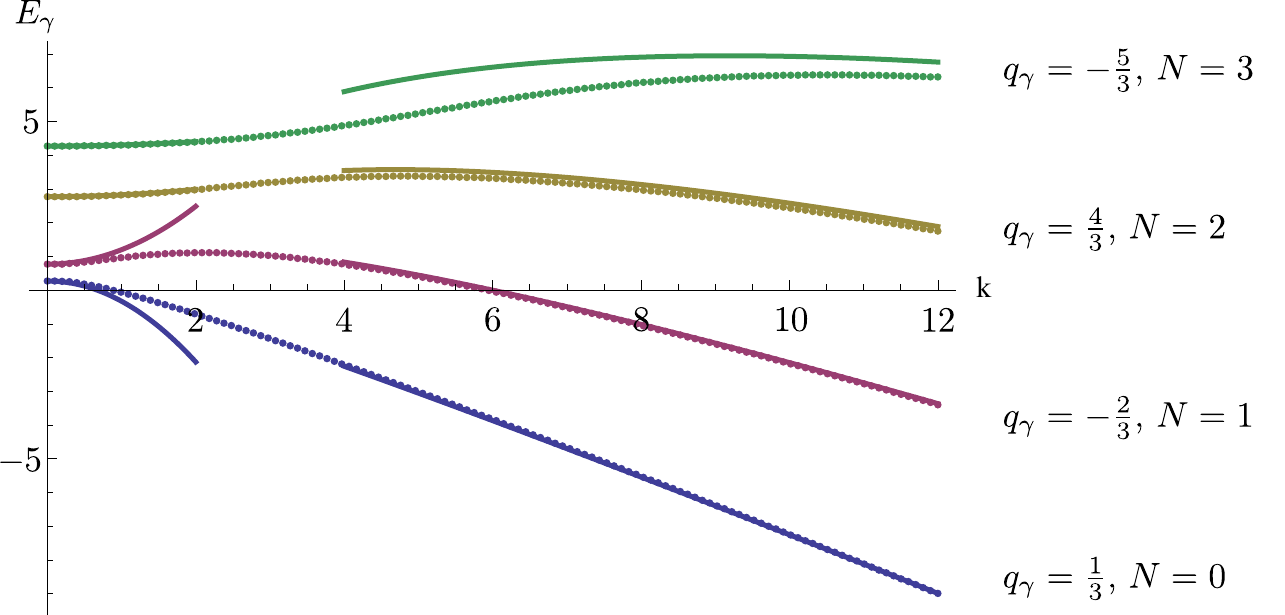}
	\caption{The energy spectrum for $(\mathcal{I},\mathcal{J}) = (0,1)$ and $d=1$ as $k$ varies. As in figure \ref{fig:pereng}, these values give $q_\gamma \equiv \frac{1}{3} (\text{mod } 1)$. Our analytic expressions are represented by the bold lines while numerical results are displayed as dots. The $N^\text{th}$ excited free state (and thus the free state in the $N^\text{th}$ band) flows to the $N^\text{th}$ excited state of the tight binding limit.}
	\label{fig:engnums}
\end{figure}

Due to the lattice structure, the labels we used for the free states continue to label the states in the tight binding limit. Since there is no crossing for fixed $\mathcal{I}$ and $\mathcal{J}$, the $N^\text{th}$ excited free state (which has the label $q_\gamma$ where $q_\gamma \in [-\frac{N+1}{2},-\frac{N}{2}]\cup[\frac{N}{2},\frac{N+1}{2}]$) flows smoothly to the state labelled by $N$ in the tight binding limit. This is confirmed by numerical calculations as seen in figure \ref{fig:engnums}, which shows the analytic and numerical energies as a function of $k$ for the four lowest energy states for fixed $(\mathcal{I}, \mathcal{J}) = (0,1)$. The eigenvalues $E_\gamma$ are found using a shooting method.

From the numerical data in figure $6$ we see that the analytic expressions \eqref{perteng} and \eqref{endo} have different regions of validity depending on which state we examine. We can explain this as follows. The large $k$ calculation relied on two approximations: that the wavefunction is concentrated within a unit cell and that it is concentrated within a region where we may expand the potential to quartic order. If we satisfy the second constraint we certainly satisfy the first so we shall examine the second. The expansion \eqref{pot} is, very roughly, good for $|\gamma| < 2$. Thus, we need the wavefunction to be decaying exponentially there. For large $\gamma$, $\phi_n$ is of the form
\begin{align}
\phi_N &\sim \gamma^N\kappa^\frac{N}{4}\exp\left(-\frac{\gamma^2}{2}\kappa^\frac{1}{2}\right) \nonumber  \\
&= \exp\left( N\log\gamma + \frac{N}{4}\log\kappa - \frac{\gamma^2\kappa^{\frac{1}{2}}}{2}\right)
\end{align}
where we have defined $\kappa = \frac{kd^2I}{1+2d^2}$. For the wavefunction to be concentrated within $-2 < \gamma < 2$ we require
\begin{equation}
N\log2 + \frac{N}{4}\log\kappa - 2\kappa^{\frac{1}{2}} < -c
\end{equation}
where $c$ is some positive constant. We see that as $N$ increases we need a larger $\kappa$, and hence $kI$, for our approximation to be valid, as the numerical results confirm. 

The regions of validity of the small $k$ perturbative energy expansion \eqref{perteng} can be explained by calculating the next non-trivial term. It is
\begin{equation}
k^4\frac{(d^2I)^3}{(1+2d^2)^3}\frac{20q_\gamma^2 + 7}{(4q_\gamma^2-1)^3(4q_\gamma^2-4)} \,.
\end{equation}
Away from degenerate points, this goes as $k^4q_\gamma^{-6} $ and as such is small for states with large $q_\gamma$. This explains why the perturbative energy calculation works for a larger range of $k$ for states with larger $q_\gamma$. 

\begin{figure}[ht]
	\centering
	\includegraphics[width=14cm]{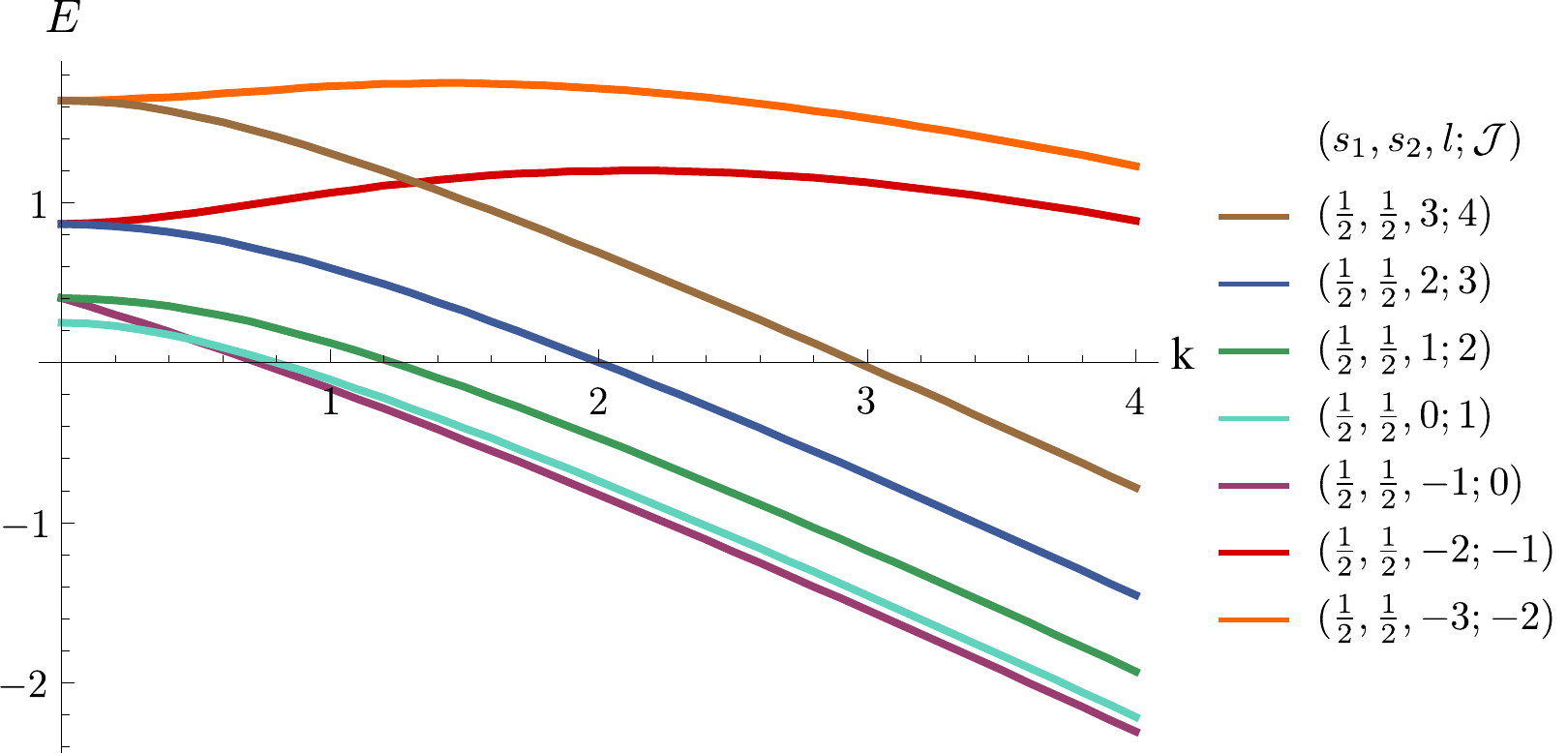}
	\caption{The energy spectrum for some low lying states with various values of $\mathcal{J}$, with $\mathcal{I} = 0$ and $s_1 = \frac{1}{2}$. Each is labelled by their $(s_1,s_2,l;\mathcal{J})$ value at $k=0$. In all but the extreme case, $l=-1$, the energetically favoured states have spin and orbital angular momentum aligned.}
	\label{fig:fullspec}
\end{figure}

The problem has now been solved in both small and large $k$ limits. Thanks to the lattice structure we can extrapolate the free states to the large $k$ states without fear of crossing between states. We have found that the energetically favourable states in the large $k$ limit come from free states which satisfy
\begin{equation} \label{ineq}
|q_\gamma| = \frac{1}{2+4d^2}\left| l - 2d^2\left(s_1+s_2\right)\right)| \leq \frac{1}{2}\,.
\end{equation}
This is our form of spin-orbit coupling. States with orbital angular momentum and spins aligned are more likely to satisfy the inequality while they are less likely to if they are anti-aligned. Note the connection between the classical minimum energy condition \eqref{cog} and our energetically favourable state condition \eqref{ineq}. To be more definite let us fix the spins of the discs be $\pm\frac{1}{2}$. Then $s_1 + s_2$ can be $1,0$ or $-1$. Take $s_1+s_2=1$ first. Energetically favoured states satisfy
\begin{equation}
|l - 2d^2| \leq 1+2d^2\,.
\end{equation} 
Thus these states have orbital angular momentum $l \in [-1,1+4d^2]$. As $l$ is usually positive, spin and orbital angular momentum are usually aligned. The extreme case, $l = -1$, corresponds to a degenerate point in the energy spectrum. Here, our labels lose meaning and we cannot distinguish between the free states $(s_1,s_2,l) = (\frac{1}{2},\frac{1}{2},-1)$ and $(s_1,s_2,l) = (-\frac{1}{2},-\frac{1}{2},1)$ as the potential is turned on. Both of these have spin and orbital angular momentum anti-aligned. The result is essentially the same for $s_1+s_2 = -1$. Here $l$ is always non-positive, except for the degenerate state. As this example demonstrates, the direction of the spins is correlated with the direction of the orbital angular momentum for most of the energetically favoured states. When $s_1+s_2 = 0$ the condition \eqref{ineq} reduces to
\begin{equation}
|l| \leq 1+2d^2\,.
\end{equation}  
This time there is no spin-orbit coupling as the orbital angular momentum has no preferred direction. Figure \ref{fig:fullspec} displays how the energy spectrum changes as $k$ is turned on for the lowest energy states which satisfy $\mathcal{I} = 0$ and $s_1 = \frac{1}{2}$. We focus on these states as this is where the spin-orbit force is present in our model. Note that states with equal $|l|$ in the free case become non-degenerate for positive $k$, just as they do in traditional spin-orbit coupling. For this figure, we take $I = 1$, $d=0.9$, with $d$ not equal to $1$ so that we avoid certain degeneracies.

In the large $k$ limit only those states which came from free states with $|q_\gamma| \leq \frac{1}{2}$ are contenders for the ground state. These are then ordered by the $\mathcal{I},\mathcal{J}$ energy contribution. This limit is exactly rigid body quantisation and in the strict limit the wavefunction is a delta function, the system completely fixed in the attractive channel. Physically, we expect the true strength of $k$ to be between the two limits we understand analytically. This is also seen in the traditional spin-orbit force: the coupling is strong enough that it has an effect on the energy spectrum but weak enough that an understanding of the spectrum without the force is vital too.

We may do an analogous calculation for the sliding configuration from figure \ref{fig:2Dslice}c. The calculation is very similar to the one above and the main physical consequence is that the energetically favoured states come from free states with small $s_1 - s_2$. Thus, the sliding configuration couples the spins. This is what is required for the tensor force -- another key ingredient in nucleon-nucleon interactions. Thus, our model unifies the spin-orbit force and the tensor force while giving them both a classical microscopic origin. In the full $3$D model both sliding and rolling motion can occur simultaneously and both need to be taken into account at the same time.

\subsection{Unequal discs}

\begin{figure}[ht]
	\centering
	\includegraphics[width=9cm]{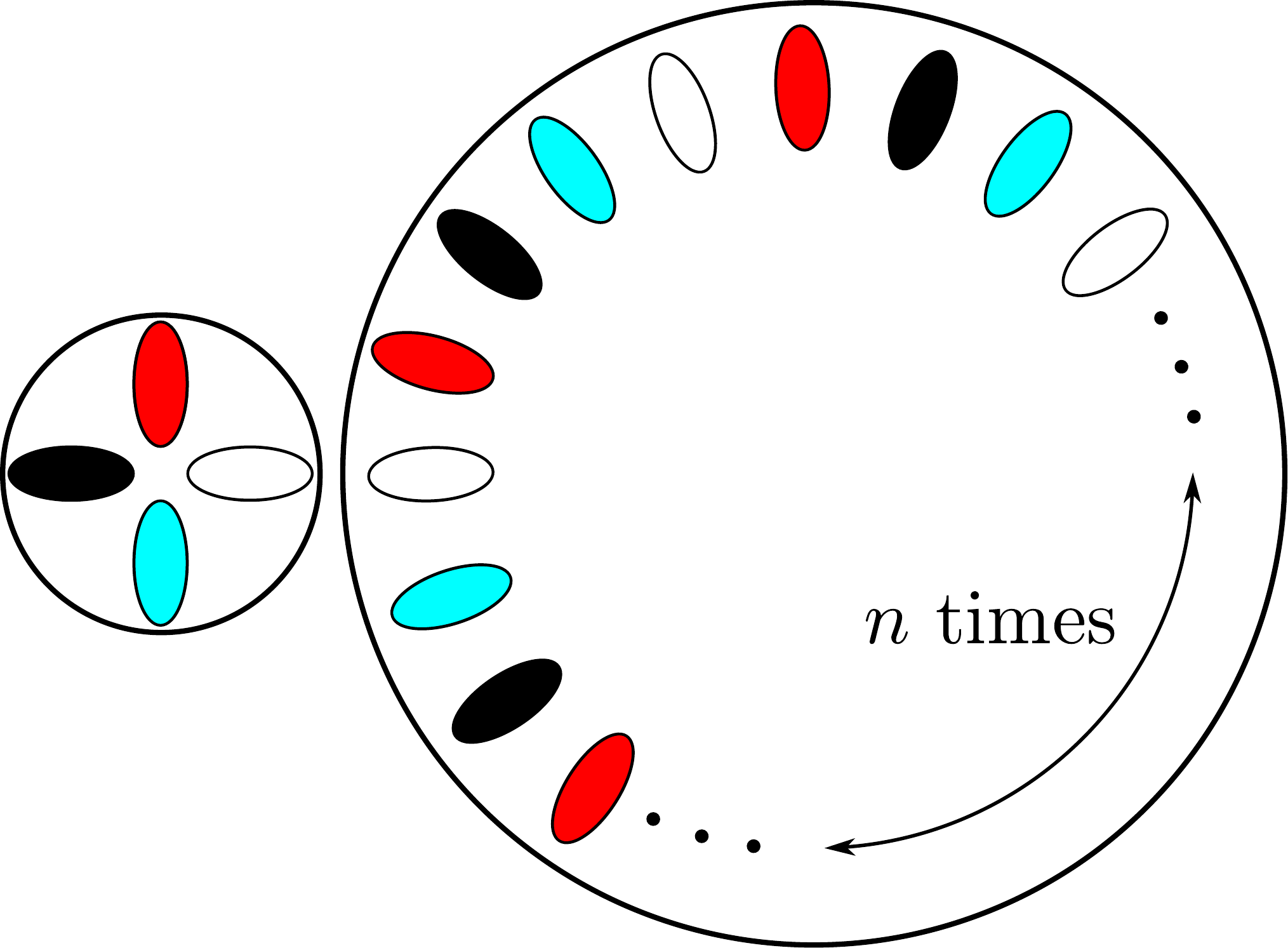}
	\caption{A small disc orbiting a large disc.}
	\label{fig:lilbig}
\end{figure}

Consider a generalisation of the system. Now a small disc orbits a larger one as seen in figure \ref{fig:lilbig}, with small and large discs labelled $1$ and $2$ respectively. Let the colour field repeat $n$ times along the edge of the large disc. This is a model for a nucleus with baryon number one more than a magic number, with a single nucleon orbiting a core. The core is generally a boson and this is how we treat the large disc. Defining our variables analogously to the variables in the previous section and using the initial configuration as in figure \ref{fig:lilbig}, the Lagrangian \eqref{lag} is modified to
\begin{equation}
L = \frac{1}{2}I_1\dot{\alpha}_1^2 + \frac{1}{2}I_2\dot{\alpha}_2^2 + \frac{1}{2}\mu r^2\dot{\beta}^2+ k\cos\left((n+1)\beta - \alpha_1 - n\alpha_2\right)\,.
\end{equation}

The classical conserved quantities are now
\begin{align}
\mathcal{J} &= I_1\dot{\alpha_1} + I_2\dot{\alpha_2} + \mu r^2\dot{\beta} = s_1+s_2+l\,, \\
\mathcal{I} &= nI_1\dot{\alpha_1} - I_2\dot{\alpha_2}=ns_1 - s_2 \,.
\end{align}
We are using the same notation as before: $l$ is the orbital angular momentum while $s_i$ is the spin of disc $i$. The classical minimum energy solution is when the discs are locked in the attractive channel and thus act like cogwheels. This gives a condition on the momenta of the system as follows:
\begin{equation} \label{classact}
I_1I_2(n+1)l - I_2\mu r^2s_1 - I_1\mu r^2ns_2 = 0\,.
\end{equation}

We can change coordinates so that the potential depends on one angle, $\gamma$, while the others are conjugate to $\mathcal{J}$ and $\mathcal{I}$. Further, we can insist that the Hamiltonian splits into two independent sectors (one depending only on $\mathcal{J}$ and $\mathcal{I}$, the other determined purely by the $\gamma$ sector) as we did in the previous section. Once again, this gives a unique coordinate transformation
\begin{equation} \label{changeo2}
\begin{pmatrix} \gamma \\ \xi \\ \eta \end{pmatrix} = \begin{pmatrix} -1 & -n & n+1 \\ \frac{I_1I_2(n+1)}{C}  & \frac{I_1I_2n(n+1)}{C} & \frac{\mu r^2(I_2+I_1n^2)}{C} \\ \frac{I_1(I_2 + I_2n + \mu r^2n)}{C} & -\frac{I_2(I_1+I_1n+\mu r^2)}{C} & \frac{\mu r^2(I_2-I_1n)}{C} \end{pmatrix} \begin{pmatrix} \alpha_1 \\ \alpha_2 \\ \beta \end{pmatrix}
\end{equation}
where $C = I_1I_2(n+1)^2 + I_1\mu r^2 n^2+ I_2\mu r^2$. This, combined with Bloch's theorem gives us the form of the wavefunction after canonical quantisation. It is
\begin{equation}
\psi(\gamma,\eta,\xi) = e^{i\mathcal{J}\xi}e^{i\mathcal{I}\eta}e^{iq_\gamma\gamma}\tilde{w}(\gamma)\,.
\end{equation}
where $\tilde{w}$ has period $2\pi$ and $q_\gamma = (I_1I_2(n+1)l - I_2\mu r^2s_1 - I_1\mu r^2ns_2)C^{-1} $. Since the small disc is a fermion, $s_1$ must be a half-integer while $s_2$ and $l$ are both integers.  Once again, the allowed values of $q_\gamma$ are separated by an integer. The Schr\"odinger equation is now
\begin{equation}
-\frac{C}{2I_1I_2\mu r^2}\frac{d^2}{d\gamma^2}\left(e^{iq_\gamma\gamma}\tilde{w}\right) - k\cos\gamma \, e^{iq_\gamma\gamma}\tilde{w} = E_\gamma e^{iq_\gamma\gamma}\tilde{w}\,,
\end{equation}
where the energy of the system is
\begin{align}
E &= E_\gamma + \frac{I_1n^2 + I_2}{2C}\mathcal{J}^2 + \frac{I_2-I_1n}{C}\mathcal{I}\mathcal{J} + \frac{I_1+I_2+\mu r^2}{2C}\mathcal{I}^2\\
&\equiv E_\gamma + E_{\mathcal{I},\mathcal{J}}\,.
\end{align}
This is simply equation \eqref{schro} with an adjusted mass. Thus we may apply all our analysis from the previous section to this problem; namely we can reuse the equations \eqref{perteng} to \eqref{endo} with the replacement
\begin{equation}
\frac{1+2d^2}{d^2I} \to \frac{C}{I_1I_2\mu r^2}\,.
\end{equation}

The physical consequence is that when $k$ is increased, the energetically favourable states have small
\begin{equation}
q_\gamma = \frac{1}{ I_1I_2(n+1)^2 + I_1\mu r^2 n^2+ I_2\mu r^2}\left( I_1I_2(n+1)l - I_2\mu r^2s_1 - I_1\mu r^2ns_2 \right)\,.
\end{equation}
Note the relationship between this and the classical condition \eqref{classact}.

To gain more insight we must estimate the moments of inertia. First we assume that the circumference of the large disc is $n$ times the circumference of the small one. Then we use the Skyrmion inspired approximation that the radius of solutions with baryon number $B$ scales as $B^{\frac{1}{3}}$ and that their mass scales linearly with $B$. Finally, we assume that the discs are touching. These give us $I_2$ and $\mu r^2$ in terms of $I_1 $ as follows:
\begin{equation}
I_2 = n^5I_1 \,, \quad \mu r^2 = \frac{2n^3}{n^3 + 1}(n+1)^2I_1\,.
\end{equation} 
It follows that
\begin{equation}
\frac{C}{I_1I_2\mu r^2} = \frac{3}{2I_1}\,\frac{n^3+1}{n^3}
\end{equation}
and
\begin{equation}
q_\gamma = \frac{1}{3(n+1)}\left( l - \frac{2n^3}{n^2-n+1}s_1 - \frac{2}{n(n^2-n+1)}s_2\right)\,.
\end{equation}
In \eqref{classact} we saw that the classical minimum energy solution obeyed $q_\gamma = 0$. If we also demand that the core is inert ($s_2=0$) then $l$ scales as $2ns_1$ for large $n$. This gives a natural explanation why orbital angular momentum increases as the size of the core increases, a relationship obeyed by the first few magic nuclei.
 
We also see that if $s_2$ is non-zero, its contribution does not have much effect on the value of $q_\gamma$; the most important contribution is from the first two terms. Naively this looks promising: after quantisation, energetically favoured states obey $|q_\gamma| \leq \frac{1}{2}$ and this can be achieved by having $s_1$ and $l$ aligned. However, the number of energetically favoured states is rather large. To be concrete, let us fix $s_1 = \frac{1}{2}$ and $s_2 = 0$ from now on. Then
\begin{equation}
q_\gamma = \frac{1}{3(n+1)}\left(l - \frac{n^3}{n^2-n+1}\right)\,.
\end{equation}
To satisfy $|q_\gamma| \leq \frac{1}{2}$ we require
\begin{equation} \label{lrange}
l \in \left[ -\frac{n^3+3}{2(n^2-n+1)},\frac{5n^3+3}{2(n^2-n+1)} \right] \,.
\end{equation}
Thus, the restriction to energetically favourable states is in fact not very limiting and the range of allowed values of $l$ grows with $n$. The centre of this range corresponds to the classical minimum energy solution, $q_\gamma = 0$. In the $k=0$ limit the states are ordered by $|l|$. As $k$ increases we become more interested in the energetically favoured states. These are ordered, in the extreme large $k$ limit, by $E_{\mathcal{I},\mathcal{J}}$. In terms of $l$ this quantity is
\begin{align}
E_{\mathcal{I},\mathcal{J}} &= \frac{1}{24I_1n^3(n^3+1)}\left( 4l^2\left(n^2-n+1\right)^2 + 4ln^3\left(n^2-n+1\right) + n^3\left(n^3+3\right)\right) \nonumber \\
&= \frac{1}{24I_1n^3(n^3+1)}\left(4(n^2-n+1)^2\left(l + \frac{n^3}{2(n^2-n+1)}\right)^2 + 3n^3\right)\,.
\end{align}
This means that the states are ordered energetically by the magnitude of $|l +\frac{n^3}{2(n^2-n+1)}|$. From comparison with \eqref{lrange} we see that the state with minimal $E_{\mathcal{I},\mathcal{J}}$ lies within the energetically favoured range of $l$ values. Thus the ground state of the system in the large $k$ limit has spin and orbital angular momentum anti-aligned as $l$ is negative, going against our classical intuition.

\begin{figure}[hr]
	\centering
	\includegraphics[width=14cm]{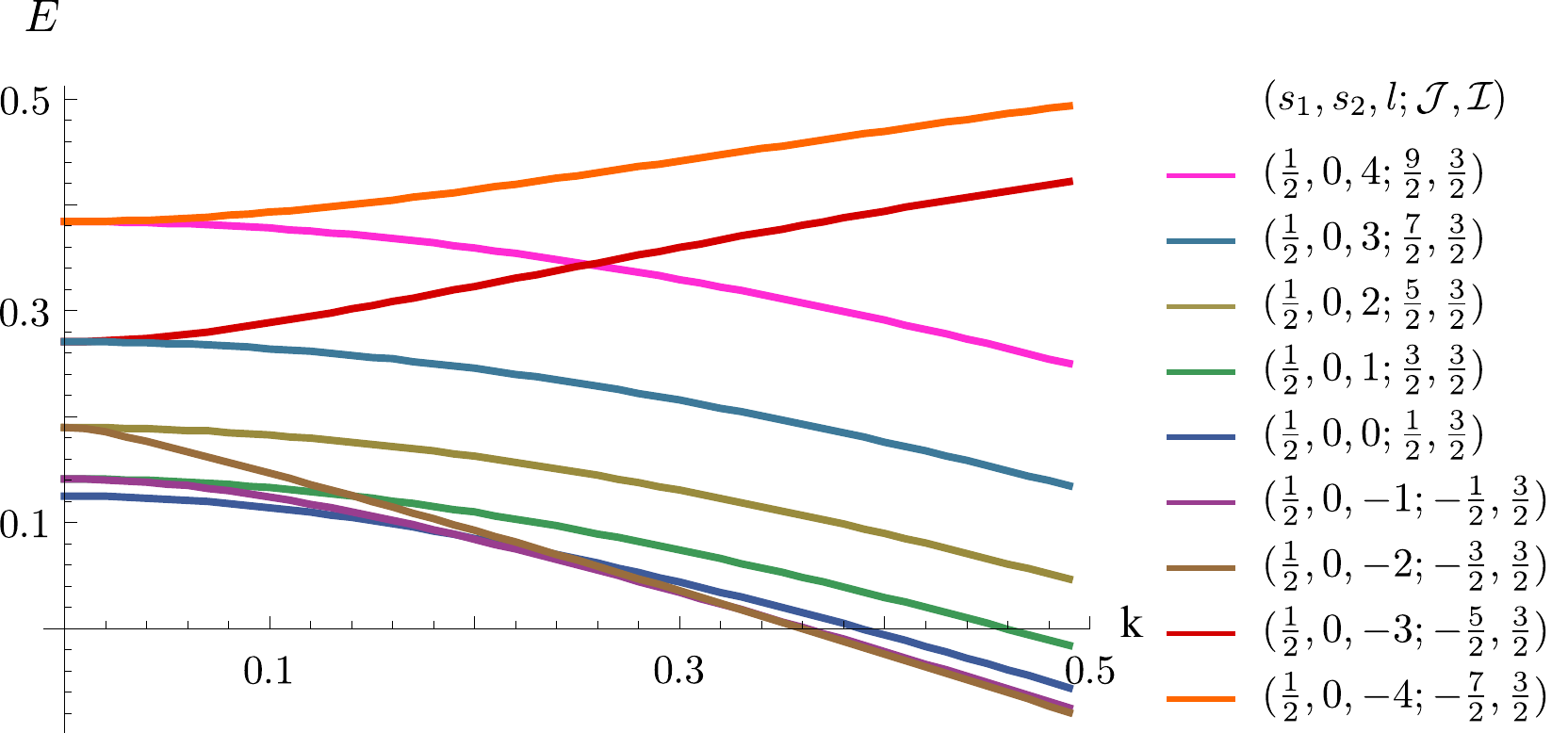}
	\caption{Energy for a variety of low lying states of unequal discs with $n=3$, as a function of $k$. Here all states with $s_1 =\frac{1}{2}$, $s_2 = 0$ and $l \in [-4,4]$ are shown. For large $|l|$ the states with $s_1$ and $l$ aligned are favoured. However for small $|l|$, the opposite is true.}
	\label{fig:lilbig2}
\end{figure}

Let us consider $n=3$ in detail to illustrate these points more concretely. Here, there are twelve energetically favoured states, with $l \in [-2,9]$. Two of these have $l$ and $s_1$ anti-aligned and these two are the lowest energy states in the large $k$ limit. However, most of the energetically favoured states do have spin and orbital angular momentum aligned. The energy, as a function of $k$, of the states with $l \in [-4,4]$ is plotted in figure \ref{fig:lilbig2}.

\pagebreak

\section{Conclusions}

The Skyrme model provides a classical microscopic origin for the spin-orbit force based on the classical pion field structure. In this paper, we have constructed a model of interacting Skyrmions based on discs interacting through a contact potential which depends only on their relative colouring. The classical behaviour resembles a pair of cogwheels and our quantisation of the model has shown that most low energy states have their spin and orbital angular momentum aligned. However, the ground state does not.

To make any real predictions from the model we must extend it to three dimensions. This is considerably more difficult as there will be three relative orientations on which the potential depends, instead of one. There is also work to be done in the Skyrme model itself. Dynamical solutions of the model which look like a $B=1$ Skyrmion orbiting a core have not yet been found.

\section*{Acknowledgements}

C. J. Halcrow is supported by an STFC studentship.

\end{document}